\documentclass[english,10pt, notitlepage, superscriptaddress]{revtex4}
\usepackage[T1]{fontenc}
\usepackage[latin9]{inputenc}
\usepackage{amstext}
\usepackage{graphicx}

\makeatletter
\@ifundefined{textcolor}{}
{%
 \definecolor{BLACK}{gray}{0}
 \definecolor{WHITE}{gray}{1}
 \definecolor{RED}{rgb}{1,0,0}
 \definecolor{GREEN}{rgb}{0,1,0}
 \definecolor{BLUE}{rgb}{0,0,1}
 \definecolor{CYAN}{cmyk}{1,0,0,0}
 \definecolor{MAGENTA}{cmyk}{0,1,0,0}
 \definecolor{YELLOW}{cmyk}{0,0,1,0}
 }

\usepackage{babel}
\def\be{\begin{equation}}
\def\ee{\end{equation}}
\def\bea{\begin{eqnarray}}
\def\eea{\end{eqnarray}}

\def\f{\frac}
\def\l{\left}
\def\r{\right}
\def\d{{\rm d}}

\makeatother

\usepackage{babel}
\begin{document}

\title{Cosmological constraints for an Eddington-Born-Infeld field}

\author{Antonio De Felice}

\affiliation{ThEP's CRL, NEP, The Institute for Fundamental Study, Naresuan University,
Phitsanulok 65000, Thailand}

\affiliation{Thailand Center of Excellence in Physics, Ministry of Education,
Bangkok 10400, Thailand}

\author{Burin Gumjudpai}

\affiliation{ThEP's CRL, NEP, The Institute for Fundamental Study, Naresuan University,
Phitsanulok 65000, Thailand}

\affiliation{Thailand Center of Excellence in Physics, Ministry of Education,
Bangkok 10400, Thailand}

\author{Sanjay Jhingan}

\affiliation{Centre for Theoretical Physics, Jamia Millia Islamia, New Delhi 110025,
India}

\date{\today}
\begin{abstract}
We consider the Eddington-Born-Infeld (EBI) model here without assuming any cosmological constant. The EBI scalar field is 
supposed to play a role of  both dark matter and dark energy. Different eras in cosmology are 
reconstructed for the model. A comparison is drawn with $\Lambda$CDM model using Supernova Ia, WMAP7 and BAO data. It seems 
that the EBI field in this form does not give good fit to observational data in comparison to the $\Lambda$CDM model.
\end{abstract}
\maketitle

\section{Introduction}
The cosmic acceleration is now considered to be one of the frontier quest of fundamental physics. Confirmed by observations (\cite{Riess:1998cb} and \cite{Scranton:2003in}), understanding of acceleration is yet to be satisfied in the regime of standard general relativity. Attempts to explain the acceleration could be performed by adding extra components of fluid into energy-momentum tensor part of the Einstein field equation. This extra component, dubbed dark energy, gives negative pressure so that it is able to drive the acceleration, see e.g. references in \cite{padma04} for reviews.  At smaller scales, a problem of an extra attractive gravity in galaxies and galaxy clusters shows up.  Effects of extra gravity such as flat galactic rotational curve, gravitational lensing, bulk velocity and structure formation are explained with dark matter \cite{Zwicky33}. On observation side, the simplest model of dark matter and dark energy-the $\Lambda$CDM model is favored however suffering from fine-tuning problem. At present, nature of dark sectors is still unknown. There is another more radical way of acquiring acceleration. That is to modify the gravity term in the action (see \cite{DeFelice:2010aj} for recent reviews).

It is possible that dark energy and dark matter is only a single entity and it could effectively have different behaviors at early and late times. This unified scenario is considered  in Chaplygin gas model \cite{Kamenshchik:2001cp}. The Chaplygin gas model however can not satisfy cosmic microwave background power spectrum and structure formation \cite{Sandvik:2002jz}.  Another idea of unifying dark sectors was proposed recently by Banados \cite{Banados:2007qz} and \cite{Banados:2008rm}. The
model  called Eddington-Born-Infeld (EBI) gravity can account for both dark matter and dark energy components without additional degrees of freedom in energy-momentum tensor. In the model, Einstein gravity couples to Born-Infeld theory giving rise to a bi-metric theory. The second metric of the theory is generated from the Born-Infeld Christoffel symbol,
$C^{\rho}_{\mu\nu}$ which is solely responsible for dark sectors. The theory predicts a dust-like effective equation of state at large scale, while at late time it behaves like a cosmological constant. The theory can also accommodate flat galactic disk-rotational curves. The model is  motivated by a combination of Eddington's idea of purely affine theory of gravity without using metric \cite{Eddington}, Born-Infeld-Einstein action \cite{Deser:1998rj}, and the idea of magnetic spin symmetry breaking in presence of an external magnetic field. Considering that topological manifold is invariant under full diffeomorphism group of transformation,
Riemannian manifolds are invariant under smaller class of subgroup of metric isometries. The Eddington action which is diffeomorphism invariant hence is considered as an unbroken state theory \cite{Witten1}. Moreover, it is  a ghost-free theory.  Introducing  $g_{\mu \nu} \neq 0$ to the gravitational theory would break this symmetry similar to having external magnetic field applying to a random spin system. The external magnetic field also breaks the symmetry of the spin system. If we let the metric couple to the Eddington action, the result is the EBI action. In the action, there is the Einstein-Hilbert part and the EBI part (see \cite{Banados:2007qz, Banados:2008rm, Banados:2008jx} for more detailed discussion).
 In context of anisotropic universe with Bianchi type I model, at late time, the EBI gravity effectively behaves like Einstein-Hilbert cosmology plus cosmological constant. The EBI term is stable at dark matter phase but also gives rise to anisotropic pressure and the perturbation decays oscillatory in time which differs from standard exponential decay case \cite{Rodrigues:2008kv}. Considering the dark energy phase, the Born-Infeld as dark energy is not stable and it produces a very strong Integrated Sachs-Wolfe effect on large scales. This suggests that for the model to be viable, cosmological constant is needed in the action \cite{Skordis}.
 When adding a cosmological constant into the model, the model still predicts too large CMB fluctuation compared to WMAP5 data. However while restricting the EBI field as dark matter, the EBI model is a best fit with the $\Lambda$CDM prediction \cite{Banados:2008fj}. The idea that Eddington action is a starting point for general relativity is pursued further more  when considering a Born-Infeld part and a cosmological constant but without having Einstein-Hilbert term. The idea was investigated in Palatini formulation to include matter fields. For homogeneous and isotropic space-time, the model present a non-singular cosmology at early time as well as non-singular collapsing of compact objects \cite{Banados:2010ix, Pani:2011mg, Delsate:2012ky, EscamillaRivera:2012vz}. In such a scenario, Poisson equation is modified and Jean length is equal to the fundamental length of the theory. Also the critical mass for a blackhole to form is equal to the fundamental mass of the theory \cite{Avelino:2012ge, Pani:2012qb}.

Here in this paper, we consider EBI model without cosmological constant as originally proposed in \cite{Banados:2008rm}. In fact, we believe that introducing a cosmological constant would make this model less attractive, as the model was introduced as a way to explain both dark energy and dark matter at the same
time. In other words, adding a  cosmological constant by hand
would mean that the model achieves only half of the original goals it was introduced for. We re-emphasize the inviability of the original EBI  model by fitting it with WMAP7, BAO and Supernova type Ia data. Compared to the study of \cite{Banados:2008fj}, where the authors studied the growth of structure for these models (they studied the evolution of the cosmological perturbations during radiation and matter domination) we perform a study of the background and look for the constraints on it coming from the most recent data.
In section II, we briefly describe the EBI model as a bi-metric theory and its cosmology. The equations of motion are described in section III. We consider cosmological era in section IV and numerical result are shown in section V. We conclude in section VI.

\section{Eddington-Born-Infeld cosmology}

 In the EBI model studied here the action has three variables, the metric $g_{\mu\nu}$, the Born-Infeld connection $C^{\rho}{}_{\mu \nu }$ and the matter field $\Psi$.   The EBI action is
\be
S[g_{\mu\nu}, C^{\rho}{}_{\mu \nu}, \Psi]   \;=\;  \frac{1}{16 \pi G} \int \d^4 x \l( \sqrt{| g_{\mu \nu} |}\, R \,+\, \frac{2}{\alpha l^2}  \sqrt{| g_{\mu\nu}  - l^2 K_{\mu \nu}  |}   \r) \;+\;  \int  \d^4 x \, {\mathcal{L}}_{\rm m}\l( \Psi, g_{\mu \nu} \r)  \label{EBIaction}  \ee
The action above has two extra constants, the length scale $l$ which resembles the dimension of a length or $1/\sqrt{R}$, and $\alpha$, which is a dimensionless parameter. The Born-Infeld Ricci tensor $K_{\mu \nu}$ is symmetric under interchanging $\mu$ and $\nu$ which is a result of symmetric properties of the Born-Infeld connection $C^{\rho}{}_{\mu \nu }$. As in standard general relativity,
\be K_{\mu \nu}  \;\equiv \;  K^{\rho}{}_{\mu \rho \nu}   \ee
where
\be  K^{\rho}{}_{\mu \alpha \nu} \; = \; C^{\rho}{}_{\mu  \nu   , \alpha}  \,+\,  C^{\rho}{}_{\sigma  \alpha  } C^{\sigma}{}_{\mu \nu}   \,- \,  C^{\rho}{}_{\mu  \alpha, \nu}    \,- \,    C^{\rho}{}_{\sigma  \nu } C^{\sigma}{}_{\mu \alpha}. \ee
The conventional matter fields are included in the Lagrangian ${\mathcal{L}}_{\rm m}$. Since the two dynamical fields $g_{\mu \nu}$ and $C^{\rho}{}_{\mu \nu }$ are independent, the Born-Infeld connection can effectively be expressed in term of a new symmetric metric $q_{\mu \nu}(x)$,
\be C^{\rho}{}_{\mu \nu }  \;=\; \frac{1}{2} q^{\rho \sigma} \l( q_{\sigma \nu, \mu}   +  q_{\mu \sigma, \nu} -  q_{\mu \nu, \sigma}  \r)\,, \ee
giving a version of bi-metric theory. As in standard case, for
the new metric covariant derivative vanishes
\be D_{\rho} q_{\mu \nu} = 0, \ee
where the covariant derivative is performed under the Born-Infeld connection, i.e.
\be D_{\rho} q_{\mu \nu} \:\equiv \: \partial_{\rho} q_{\mu\nu} \,-\, C^{\sigma}{}_{\mu \rho}\, q_{\sigma \nu}    \,    -  \, C^{\sigma}{}_{\nu \rho }\, q_{\sigma \mu}\,. \ee
Varying the action (\ref{EBIaction}) with respect to two dynamical fields, the metric $g_{\mu\nu}$ and the connection $C^{\alpha}_{\mu\nu}$, yields the following equations of motion,
\be
G_{\mu\nu}  = \sqrt{\frac{|g_{\mu\nu}  -  l^2 K_{(\mu\nu)} |}{|g_{\mu\nu}|}}  \, g_{\mu \rho} \l(\f{1}{g - l^2 K} \r)^{\rho \sigma}  g_{\sigma \nu} +  8 \pi G T_{\mu\nu}^{\rm m}\,.  \label{fieldlong}
\ee
Defining 
\be
\sqrt{q}\, q^{\mu\nu} \, \equiv  \,  - \frac{1}{\alpha}\sqrt{|g_{\mu\nu} - l^2 K_{\mu\nu}|} \, \l( \frac{1}{g - l^2 K} \r)^{\mu\nu}      \label{ska}
\ee
hence (\ref{fieldlong}) can be written as
 \bea G_{\mu\nu}  \; &=& \; - \frac{1}{l^2}  \sqrt{\frac{|q_{\mu \nu}  |} {| g_{\mu \nu}   |}}\, g_{\mu \alpha} q^{\alpha \beta} g_{\beta \nu}  \, + \, 8 \pi G \, T^{\rm m}_{\mu \nu} \,. \label{eq_G}   \eea
Varying of the action with respect to the connection $C^{\alpha}_{\mu\nu}$, one can find $D_{\rho} (\sqrt{q} q^{\mu\nu}) = 0$
Taking determinant of (\ref{ska}) then we obtain,
 \bea K_{\mu \nu} \; &=& \;  \frac{1}{l^2} \l(  g_{\mu \nu}   + \alpha\, q_{\mu \nu}  \r)\,.  \label{eq_K} \eea 
 The first term in Eq.(\ref{eq_G}), is a modification from the Born-Infeld part. $T^{\rm m}_{\mu \nu} $ is the matter field energy-momentum tensor. These results agree with the ones first shown in \cite{Banados:2008rm}. The two metrics $g_{\mu \nu}$  and $q_{\mu \nu}$ possess homogeneity and isotropy with flat spatial curvature,
\bea  \label{met1}
g_{\mu \nu} \d x^{\mu} \d x^{\nu} &=& - \d t^2 +  a(t)^2 ( \d x^2 + \d y^2 + \d z^2), \\\label{met2}
q_{\mu\nu}  \d x^{\mu} \d x^{\nu}  &=& - X(t)^2 \d t^2 + Y (t)^2 (\d x^2 + \d y^2 + \d z^2)  \eea $ g_{tt} = -1, $ due to the gauge freedom in time. Here $X(t)$ is the time rescaling of the metric $q_{\mu \nu}$  whereas  $a(t)$ and $ Y(t)$ behave like scale factors in   $ g_{\mu \nu} $ and $ q_{\mu \nu} $ respectively. The $a(t_0)$ is set to 1 so that $H_0 = \dot{a}(t_0)$ as in \cite{Banados:2008rm}.

\section{The equations of motion}
Applying the metric ansatz Eqs. (\ref{met1}) and (\ref{met2}), to the equations of motion (\ref{eq_G})
and (\ref{eq_K}), we obtain  first order equations, which are
\begin{eqnarray}
H^{2} & = & \frac{1}{3l^{2}}\left(\frac{Y^{3}}{a^{3}}\right)\frac{1}{X}+\frac{8\pi G}{3}\,(\varrho_{m}+\varrho_{r})\,,\label{eq:fried1}\\
\frac{d}{dt}\!\left(\frac{Y^{3}}{X}\right) & = & 3XY^{3}\left(\frac{a^{2}}{Y^{2}}\right)H\,,\label{eq:sanj2}\\
\left(\frac{\dot{Y}}{Y}\right)^{2} & = & \frac{X^{2}}{3l^{2}}\left(\alpha-\frac{1}{2X^{2}}+\frac{3}{2}\,\frac{a^{2}}{Y^{2}}\right).\label{eq:sanj3}
\end{eqnarray}
It should be noted that $l$ has dimensions of length ($M^{-1}$),
whereas $Y$ has  dimensions of $a$. Finally, $X$ is dimensionless.

From Eq.~(\ref{eq:fried1}), we can introduce an energy
density as
\begin{equation}
\varrho_{X}\equiv\frac{1}{8\pi l^{2}G}\,\frac{Y^{3}}{X\, a^{3}}\,,
\end{equation}
 and by taking the derivative of the Friedmann equation, one can find
an effective pressure for this dark component as
\begin{equation}
p_{X}=w_{X}\,\varrho_{X},
\end{equation}
 where
\begin{equation}
w_{X}=\frac{1}{3}\,\frac{X'}{X}-\frac{Y'}{Y} ,
\end{equation}
and a prime denotes differentiation with respect to $N=\ln a$.
% We note here, that the fixed point in Sanjay's notes had the properties
% that $X={\rm constant}$ and $Y\propto a$. This implies that on the fixed
% point we find $w_{X}=-1$, that is the fixed point is a de Sitter one.
% Another thing we can see at this level is that if we want this dark component
% to behave as dark matter (before the de Sitter era) then we need $w_{X}=0$,
% or $Y^{3}/X={\rm constant}$, so that $\varrho_{X}a^{3}$ remains constant.

Let us now introduce the variable
\begin{equation}
\Omega_{X}\equiv\frac{8\pi G\varrho_{X}}{3H^{2}}=\frac{1}{3l^{2}}\,\frac{Y^{3}}{H^{2}\, X\, a^{3}}\,.
\end{equation}
In terms of this variable, the Friedmann equation can be written as
\begin{equation}
1=\Omega_{X}+\Omega_{m}+\Omega_{r}\,,
\end{equation}
 where we have defined, as usual,
\begin{equation}
\Omega_{m}\equiv\frac{8\pi G\varrho_{m}}{3H^{2}}\,,\qquad{\rm and}\qquad\Omega_{r}\equiv\frac{8\pi G\varrho_{r}}{3H^{2}}\,,\label{eq:rhomr}
\end{equation}
and have assumed $\rho_{m}\propto a^{-3}$, $\rho_{r}\propto a^{-4}$.

We will demand $\Omega_{X}\geq0$, as $\Omega_{X}$ represents an
effective matter density, otherwise $\Omega_{m,r}$ could assume values
larger than unity. From Eqs.~(\ref{eq:rhomr}), we find
\begin{eqnarray}
\Omega'_{m} & + & \frac{2H'}{H}\,\Omega_{m}+3\Omega_{m}=0\,,\\
\Omega'_{r} & + & \frac{2H'}{H}\,\Omega_{r}+4\Omega_{r}=0\,.
\end{eqnarray}
Eq.\ (\ref{eq:sanj2}) can then be rewritten as
\begin{equation}
\Omega'_{X}+3\Omega_{X}+\frac{2H'}{H}\,\Omega_{X}=\left(\frac{3X^{4}}{l^{4}H^{4}}\right)^{1/3}\Omega_{X}^{1/3}\,.
\end{equation}
and Eq.\ (\ref{eq:sanj3}) can  be rewritten as
\begin{equation}
H^{2}\left[1+\frac{2}{3}\frac{H'}{H}+\frac{1}{3}\left(\frac{X'}{X}+\frac{\Omega'_{X}}{\Omega_{X}}\right)\right]^{2}=\frac{1}{3l^{2}}\left[-\frac{1}{2}+\alpha X^{2}+\frac{1}{2}\left(\frac{3X^{4}}{l^{4}H^{4}}\right)^{1/3}\Omega_{X}^{-2/3}\right].
\end{equation}
 Therefore, we also need an equation for $H.$ This can be found by
differentiating the Friedmann equation as
\[
\Omega'_{X}+\Omega'_{m}+\Omega'_{r}=0\,,
\]
or
\begin{equation}
\Omega'_{X}=\frac{2H'}{H}(1-\Omega_{X})+3(1-\Omega_{X})+\Omega_{r}\,.
\end{equation}
 Therefore the dynamical autonomous equations can be written as
\begin{eqnarray}
\Omega'_{X} & = & \frac{2H'}{H}(1-\Omega_{X})+3(1-\Omega_{X})+\Omega_{r}\,,\label{eq:eom1}\\
\Omega'_{r} & = & -\frac{2H'}{H}\Omega_{r}-4\Omega_{r}\,,\\
\Omega'_{X} & = & -3\Omega_{X}-\frac{2H'}{H}\,\Omega_{X}+\left[\frac{3X^{4}}{K^{4}(H/H_{0})^{4}}\right]^{1/3}\Omega_{X}^{1/3}\,,\\
\frac{H^{2}}{H_{0}^{2}}\left[1+\frac{2}{3}\frac{H'}{H}+\frac{1}{3}\left(\frac{X'}{X}+\frac{\Omega'_{X}}{\Omega_{X}}\right)\right]^{2} & = & \frac{1}{3K^{2}}\left[\alpha X^{2}-\frac{1}{2}+\frac{1}{2}\left[\frac{3X^{4}}{K^{4}(H/H_{0})^{4}}\right]^{1/3}\Omega_{X}^{-2/3}\right],\label{eq:eom4}
\end{eqnarray}
where we have introduced the dimensionless variable $K^{2}\equiv H_{0}^{2}l^{2}$.
This shows that the present value of $H$, can be re-absorbed into the
free parameter $K$. In terms of these variables we find
\begin{equation}
w_{X}=-1-\frac{2}{3}\,\frac{H'}{H}-\frac{\Omega'_{X}}{3\Omega_{X}}\,.
\end{equation}

\section{Cosmological eras}

Let us consider the different eras in the cosmological history. We
can distinguish the following cases.
\begin{enumerate}
\item Radiation era: We can set $\Omega_{r}=1$ and $\Omega_{X}=0$. This fixes $\Omega_{m}=0$. All the equations of motion are satisfied
if
\begin{equation}
\frac{H'}{H}=-2\,,\qquad\textrm{which implies}\qquad H=\frac{1}{2t}\,,
\end{equation}
 as expected.
\item Matter era: Now we have two options:

\begin{enumerate}
\item We can assume $\Omega_{m}=1$ and  $\Omega_{r}=0$. In this case $\Omega_{X}=0$.
This implies that we are considering dark matter as an extra matter
component (inside $\rho_{m}$) and not the $X$ dark component. In
this case the equations of motion are satisfied if
\begin{equation}
\frac{H'}{H}=-\frac{3}{2}\,,\qquad\textrm{that is}\qquad H=\frac{2}{3t}\,,
\end{equation}
 as expected.
\item Now we assume that the dominant dark component behaves as dark
matter, whereas $\Omega_{m}\to\Omega_{b}$, that is, the matter component
reduces to the baryon component and we suppose it is not the dominant
one. In this case we need to impose $\Omega_{X}=1$, and $\Omega_{r}=0$.
Since we still want that $H'/H=-3/2$, the equations of motion cannot
be solved at the same time. Therefore this case shows that if the
$X$-components gives an effective dark-matter contribution in the
past, it cannot be along a fixed point solution. However, there could
be a transient solution from $\Omega_{X}=0$ and $\Omega_{X}=1$ which
could still mimic a dark-matter component.
\end{enumerate}
\item Dark energy era: In this case we set $\Omega_{m}=0=\Omega_{r}$ together
with $\Omega_{X}=1$. We look for de Sitter solution that is
$H'=0$. The equations of motion imply
\begin{equation}
\frac{X^{2}}{H_{{\rm dS}}^{2}K^{2}}=3\,.
\end{equation}
 Therefore, $X=X_{{\rm dS}}={\rm constant}$. Then for a de Sitter
solution we find
\begin{equation}
H_{{\rm dS}}=\frac{|X_{{\rm dS}}/K|}{\sqrt{3}}\,.\label{eq:Hds}
\end{equation}
 Furthermore, the equations of motion give
\begin{equation}
X_{{\rm dS}}^{2}=\frac{1}{1-\alpha}\,,
\end{equation}
 which implies that $\alpha<1$. We can also write
\begin{equation}
H_{{\rm dS}}^{2}=\frac{1}{3K^{2}(1-\alpha)}\,.
\end{equation}

\item Dark Energy for the case $\alpha>1$: Let us consider the case when,
at very late times, $X/H=\lambda\approx{\rm constant}$, $\Omega_{X}\approx1$,
$\Omega_{r}\approx0\approx\Omega_{m}$, but still $H'/H\to{\rm constant}$,
as well as $X'/X\to{\rm constant}$. Then, by neglecting any constant
term with respect to the $X$ term, we find the following two conditions
which need to be satisfied
\begin{eqnarray}
\frac{3^{1/3}}{\lambda^{4/3}K^{4/3}}-\frac{2X'}{X}-3 & = & 0\,,\\
\frac{\sqrt{\alpha}}{\sqrt{3}\lambda K}-\frac{X'}{X}-1 & = & 0\,,
\end{eqnarray}
which imply
\begin{equation}
\alpha=\frac{3\left(3^{1/3}-\lambda^{4/3}K^{4/3}\right)^{2}}{4\lambda^{2/3}K^{2/3}}\,.
\end{equation}
This solution is not a de Sitter solution, as in fact we find
\begin{equation}
w_{X}\to-\frac{1}{3^{2/3}\lambda^{4/3}K^{4/3}}\neq-1\,.
\end{equation}

\end{enumerate}

\section{Numerical discussion}

Let us consider a numerical solution of the Eqs.~(\ref{eq:eom1})-(\ref{eq:eom4}).
Rewriting  Eq.~(\ref{eq:eom4}) as
\begin{equation}
1+\frac{2}{3}\frac{H'}{H}+\frac{1}{3}\left(\frac{X'}{X}+\frac{\Omega'_{X}}{\Omega_{X}}\right)=\frac{1}{\sqrt{3}H\, l}\sqrt{\alpha X^{2}-\frac{1}{2}+\frac{1}{2}\left(\frac{3X^{4}}{l^{4}H^{4}}\right)^{1/3}\Omega_{X}^{-2/3}}\,,\label{eq:eom4b}
\end{equation}
and by allowing the constant $l$ (or $K$) to take also negative values
(but $K\neq0$), then we recover both the branches of Eq.~(\ref{eq:eom4}).
Notice that $K<0$, Eq.~(\ref{eq:Hds}), and implies $X_{{\rm dS}}<0$,
for $\alpha<1$. This further implies that the two branches, on their de Sitter
solution, will differ by the sign of the final value of $X$.

\subsection{Initial conditions}

Let us solve the equations of motion from a given redshift ($z=z_{i}\gg1$), such that at $z=z_{i}$ the universe is in the radiation era.
We will set the initial condition for the Hubble parameter, during
the radiation era, as the one given by GR, namely
\begin{equation}
H_{i}=H_{i}^{({\rm GR})}=\sqrt{\Omega_{r,0}\, e^{-4N_{i}}+\Omega_{m,0}^{({\rm GR})}e^{-3N_{i}}+(1-\Omega_{m,0}^{({\rm GR})}-\Omega_{r,0})}\,.
\end{equation}
In what follows, we will fix the value of $\Omega_{r,i}$, during
radiation era, at $N=N_{i}\equiv-\log_{10}(1+1.76\times10^{5})$,
such that $\Omega_{r}(N=0)=\Omega_{r,0}$ which will be set equal
to a fixed value. In this model, we  have five
parameters, $\Omega_{m,0}$, $X_{i}$, $\Omega_{X,i}$, $\alpha$,
$K$. However, we will fix the initial condition for $\Omega_{X,i}$
by requiring the condition $\Omega_{K,0}=1-\Omega_{m,0}-\Omega_{r,0}$
to hold. Finally, the four parameters, $\Omega_{m,0}$, $X_{i}$,
$\alpha$, $K$, will be considered to be free. In particular, since
the $X$-component is supposed to explain both dark matter and
dark energy, we will set the following
range $0< \Omega_{m,0}\leq0.4$. We run the other parameters to change
over a large range, $-200<X_{i}<200$, $-10<\alpha<10$, and $-15<K<15$.
Notice that in this parameter range, the system does not have a $\Lambda$CDM
limit, therefore one expects deviations from the concordance model.
Since this model has been introduced to explain dark energy and dark
matter at the same time, this no-$\Lambda$CDM limit is in fact
well motivated.

\subsection{Results}

We have calculated the total $\chi^2$ for this model by using WMAP7 data (the background constraints on the two CMB shift parameters \cite{WMAP7}), the BAO (SDSS7) data (two points) \cite{Percival}, and supernova type Ia (constitution data) \cite{hicken}, following the same method followed in \cite{chisq}.
The minimum for the $\chi^{2}$ is located at
\begin{equation}
\Omega_{m,0}=0.250078\,,\ K=8.629636\,,\ \alpha=2.760611\,,\ X_{i}=77.73029\,,\qquad{\rm where}\qquad\chi^{2}=\chi_{{\rm min}}^{2}=484.505\,,\label{eq:minchi}
\end{equation}
where we have also fixed $\Omega_{r,i}=0.999827$, and $\Omega_{X,i}=1.14432\times10^{-6}$
for the reasons already explained above. Trying to set priors on $\Omega_{m,0}$
like $\Omega_{m.0}=\Omega_{b,0}$ (i.e.~fixing the scalar field to
be the main source of dark matter) leads to much larger values for
$\chi^{2}$. Furthermore, data tends to prefer clearly the $\alpha>1$
case, as for $0<\alpha<1$, the $\chi^{2}$ increases.

\begin{figure}[th]
\centering\includegraphics[width=7cm]{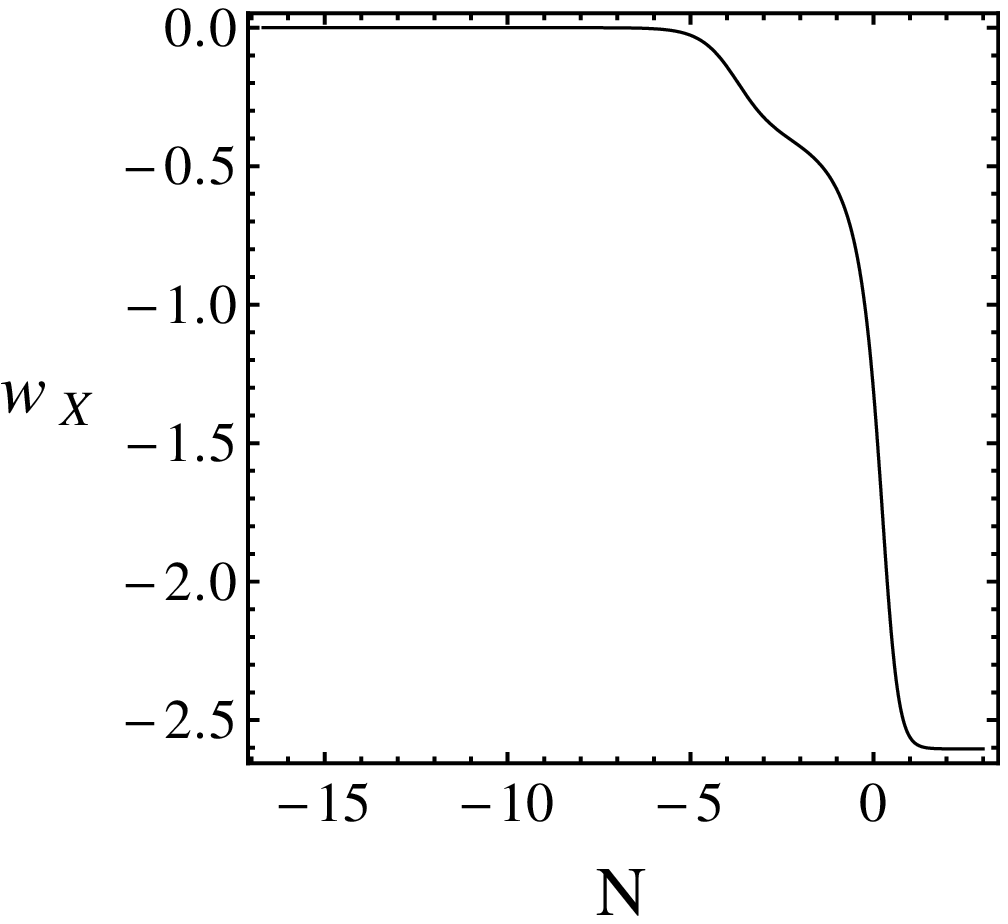}\qquad\qquad\includegraphics[width=7cm]{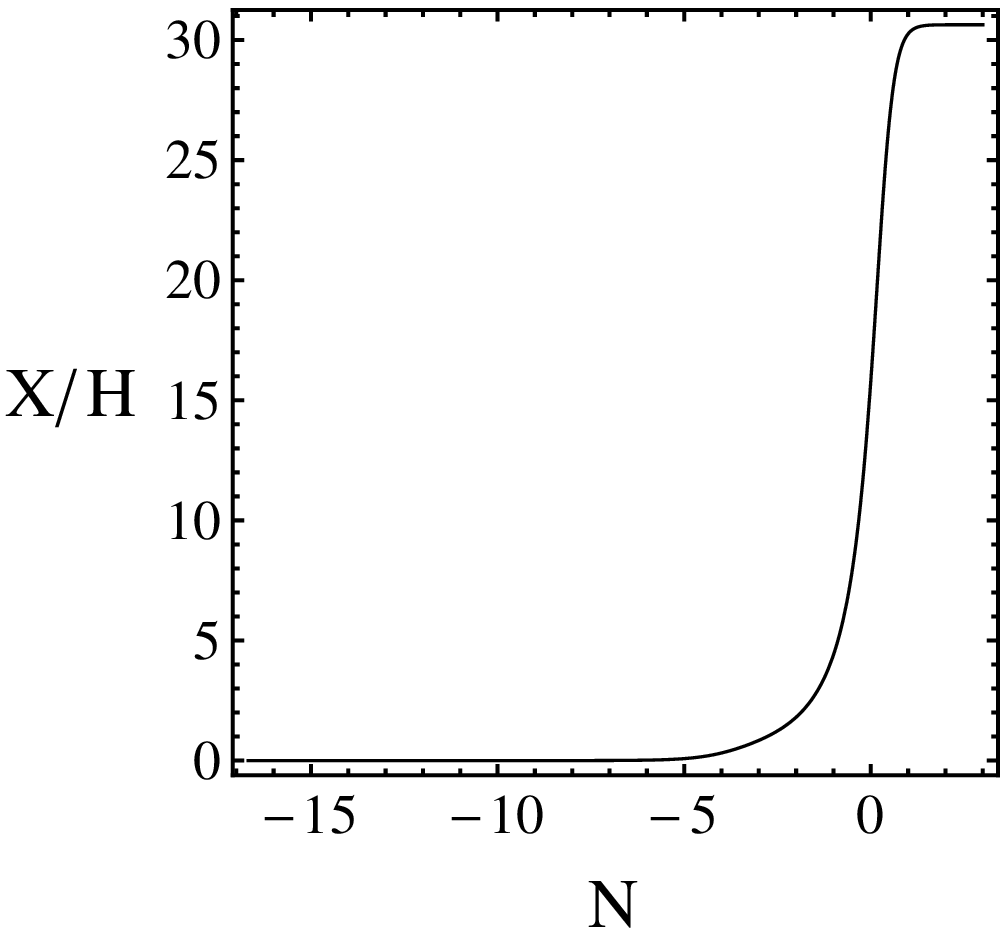}
\caption{Effective equation of state for the scalar field (left panel), and
plot of the ratio $X/H$ (right panel). At early times the Eddington-Born-Infeld
scalar field behaves as a dark matter component ($w_{X}\approx0$),
then, at late times, it drives the evolution of the universe. This
plot shows the evolution for the parameters which minimize the $\chi^{2}$
given in (\ref{eq:minchi}). Notice that since $\alpha>1$, the final
state is not de Sitter ($w_{X}<-1$), rather it tends to the solution
characterized by $H'/H\to{\rm constant}$, and $X/H\to{\rm constant}$
(right panel). \label{fig:wxmin}}
\end{figure}

Nonetheless, the minimum value for $\chi_{{\rm min}}^{2}$ is still
much larger than $\Lambda$CDM's value ($\chi_{\Lambda{\rm CDM}}^{2}\approx469$). The $\chi^2$ for $\Lambda$CDM has two free degrees of freedom ($\Omega_{m0}$, $\Lambda$) that we can vary. Therefore according to the $\chi^2$-probability distribution, at 95\% confidence level, $\Lambda$CDM rules out those models, at 2-$\sigma$, whose fit to the same data will lead to $\chi^2-\chi^2_{\Lambda{\rm CDM}}>5.99$. However, the models discussed here have $\chi^2=484.5$, so that $\chi^2-\chi^2_{\Lambda{\rm CDM}}=15.5$, which implies that these models are excluded at 2-$\sigma$.
This large difference implies the model under consideration does not
fit the data, already at 2$\sigma$, as well as $\Lambda$CDM. This
is tantamount to saying that the $\Lambda$CDM cosmological evolution rules out
this class of models. Since the $\chi^{2}$ for the model studied here
is higher than $\Lambda$CDM's one, we can deduce that data
do not support well the evolution of the effective equation of state
plotted in Fig.~\ref{fig:wxmin}. It should also be pointed out
that for the parameters for which $\chi^{2}=\chi_{{\rm min}}^{2}$,
the scalar field, although it has in the past $w_{X}\approx0$, it
is anyhow a subdominant dark matter component (since on the minimum-$\chi^2$ solution, the dust-like
dark matter contributes up to $\Omega_{m,0}\approx0.25$). This implies
that the scalar field starts dominating the evolution of the universe
only at late times, that is it contributes to the dynamics essentially
only as a dark energy field. But it is a dark energy field which,
at early times, it is quite different from a cosmological constant:
this may be part of the reason why, in this case, the model cannot not fit the data
well. One option would be adding a bare cosmological constant (as also proposed in \cite{Banados:2008fj}), but in
this case the model loses part of the interest as it would stop being an attractive
dark energy model. Furthermore the evolution tends to lead to a fast transition of the effective equation-of-state parameter. This may also contribute to a worse fit to the data compared to $\Lambda$CDM.

\begin{figure}[th]
\centering\includegraphics[width=7cm]{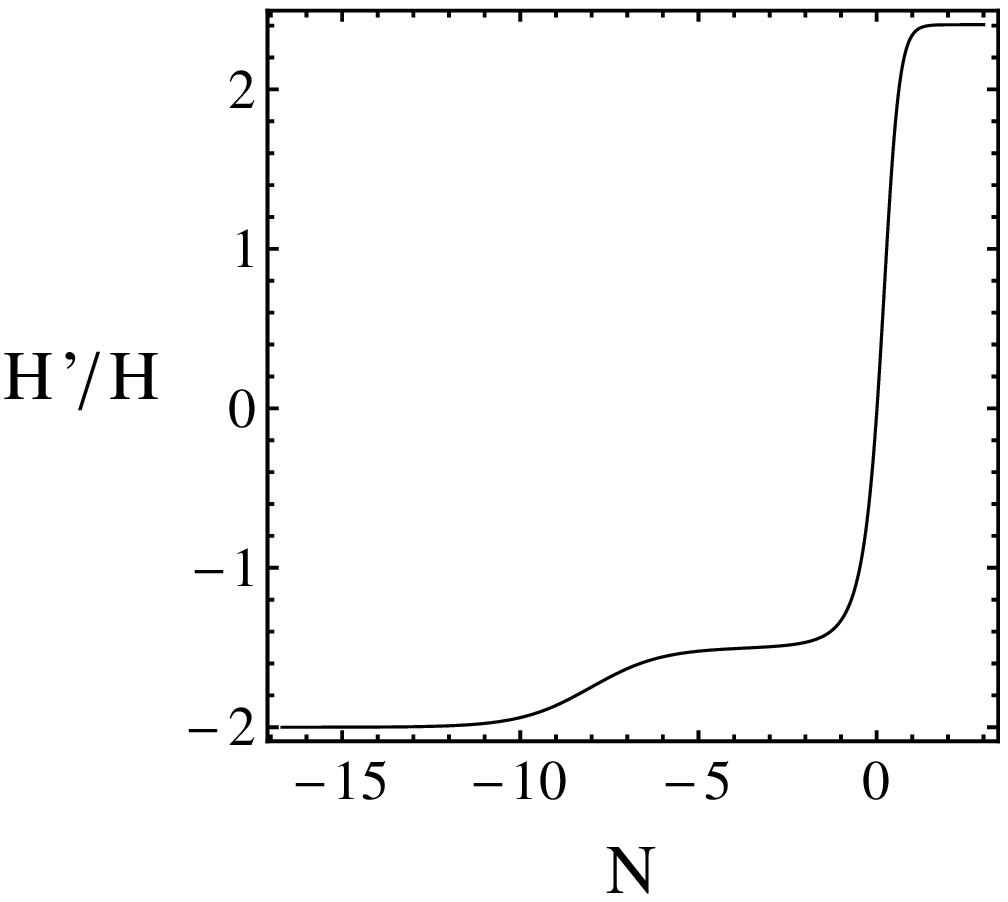}\qquad\qquad\includegraphics[width=7cm]{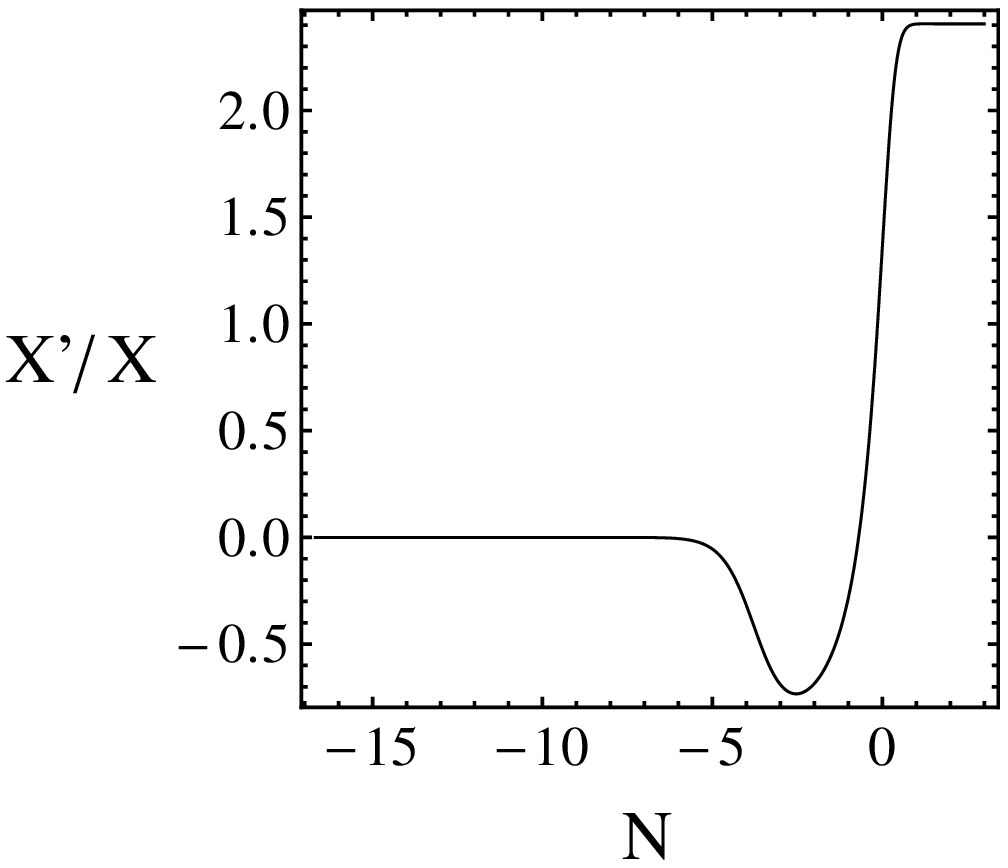}
\caption{Evolution for $H'/H=\dot{H}/H^2$ (left panel), and for the variable $X'/X$ (right panel). The evolution, starting from radiation domination, passing through matter domination, at late times, tends to a super-accelerating final state on the minimum-$\chi^2$ solution. \label{fig:wxminHH}}
\end{figure}

It should be noted that negative values for any of $X_{i}$, $\alpha$,
and $K$ leads to very large values for $\chi^{2}$ (typically larger
than 1000), giving a bad fit to the data.

\section{Conclusions}

We have studied the Eddington-Born-Infeld scalar field which was proposed
to model  both dark matter and dark energy at the same time. We have
solved the equations of motion and studied the behavior of the background
at different times: at early times, indeed the scalar field behaves
as a dark matter component with equation of state parameter $w_{X}\approx0$.
Only at late times, the field can lead the dynamics of the universe
to an accelerated regime, which depending on the parameters of the
model, is described by either a de Sitter solution, or a rather different
dynamics described by $H/X\to{\rm constant}$, and $H'/H\to{\rm constant}$
(where a prime denotes differentiation with respect to the $N=\ln a$).

The fact that, at early times, the scalar field behaves as a dust
component can in principle alleviate the problem of finding a dark
matter component, as indeed the nature of dark matter and dark energy
would have the same explanation.

In order to see whether this model is viable or not, we studied the
cosmological constraints that its dynamics have to pass when considering
WMAP7 data, BAO and Supernova type Ia. For this goal, we have calculated
the $\chi^{2}$ as a function of four free parameters, that is $K$,
$\alpha$ (two theoretical dimensionless parameters of the model)
together with $\Omega_{m,0}$ (which states how much of an extra standard
dust component is needed), and $X_{i}$, the initial value for the
time-rescaling component of the Born-Infeld metric.

We have found that the model cannot give a good fit to the data (compared
to $\Lambda$CDM), and hence the model cannot be considered viable. We have proved this statement by constraining the background. This approach differs from the one followed in \cite{Banados:2008fj}, where the authors studied the evolution of the cosmological perturbations in order to constrain the growth of structures.  In particular, we have used only the constraints on the background coming from the WMAP7 data. Instead in \cite{Banados:2008fj}, the authors considered constraints only on the perturbations power-spectrum.  It is possible, as also suggested in \cite{Banados:2008fj} that introducing a cosmological constant would improve the fit, but, on the other hand, the model would partially lose its original motivation of explaining at the same time both dark energy and dark matter.
In particular data prefers the non-de Sitter solution, preferring a
fast transition to values for $w_{X}<-1$. In this model, at early
times, a cosmological constant is absent from the beginning, as the
scalar field initially (and up to very recently) behaves as a dark
matter component.

\section*{Acknowledgments}
B. G. thanks Baojiu Li for bringing his initial attention to the EBI model. B. G. is supported by National Research Council of Thailand and the Basic Research Grant of the Thailand Research Fund.  B. G. thanks Institute for Particle Physics Phenomenology, University of Durham, U.K. for hospitality during his visit. S. J. thanks The Institute for Fundamental Study, Naresuan University, Thailand for hospitality during his visit where this work was initiated.


\begin{thebibliography}{0}


\bibitem{Riess:1998cb}
A. G. Riess {\it et al.} (Supernova Search Team Collaboration),
%Observational Evidence from Supernovae for an Accelerating Universe and a Cosmological Constant
Astron. J. {\bf 116}, 1009 (1998);
% CITATION = ASTRO-PH 9805201
S. Perlmutter {\it et al.} (Supernova Cosmology Project Collaboration),
% Measurements of Omega and Lambda from 42 High-Redshift Supernovae
Astrophys. J. {\bf 517}, 565 (1999);
% CITATION = ASTRO-PH 9812133
A. G. Riess,
% Peculiar Velocities from type Ia Supernovae
arXiv: astro-ph/9908237;
% CITATION = ASTRO-PH 9908237
G. Goldhaber {\it et al.} (The Supernova Cosmology Project
Collaboration),
% Timescale Stretch Parameterization of Type Ia Supernova B-band Light Curves
Astrophys. J., {\bf 558}, 359 (2001);
S. Masi {\it et al.},
% The BOOMERanG experiment and the curvature of the universe,
Prog. Part. Nucl. Phys. {\bf 48}, 243 (2002);
% [arXiv: astro-ph/0201137]
% CITATION = ASTRO-PH 0201137
% arXiv: astro-ph/0104382;
J. L. Tonry {\it et al.} (Supernova Search Team Collaboration),
% Cosmological Results from High-z Supernovae
Astrophys. J. {\bf 594}, 1 (2003);
% CITATION = ASTRO-PH 0305008
A. G. Riess {\it et al.} (Supernova Search Team Collaboration),
% Type Ia Supernova Discoveries at z>1 From the Hubble Space Telescope: Evidence for Past Deceleration and Constraints on Dark Energy Evolution
Astrophys. J. {\bf 607}, 665 (2004);
% CITATION = ASJOA,607,665
A. G. Riess {\it et al.},
% New Hubble Space Telescope Discoveries of Type Ia Supernovae at $z > 1$: Narrowing Constraints on the Early Behavior of Dark Energy
Astrophys. J. {\bf 659}, 98 (2007);
% CITATION = ASTRO-PH/0611572





\bibitem{Scranton:2003in}
R. Scranton {\it et al.} (SDSS Collaboration),
% Physical Evidence for Dark Energy
arXiv: astro-ph/0307335.
% CITATION = ASTRO-PH 0307335




\bibitem{padma04}
T. Padmanabhan,
% Dark Energy: the Cosmological Challenge of the Millennium
Curr. Sci. {\bf 88}, 1057 (2005);
% arXiv: astro-ph/0411044
% CITATION = ASTRO-PH 0411044
E. J. Copeland, M. Sami and S. Tsujikawa,
% Dynamics of dark energy
Int. J. Mod. Phys. D {\bf 15}, 1753 (2006);
% [arXiv: hep-th/0603057];
T. Padmanabhan,
% Dark Energy: Mystery of the Millennium
AIP Conf. Proc. {\bf 861}, 179 (2006);
% [arXiv: astro-ph/0603114]
% CITATION = APCPC,861,179
L. Amendola and S. Tsujikawa ``Dark Energy: Theory and Observations'', Cambridge University Press (2010).


\bibitem{Zwicky33}
  F. Zwicky, Helv. Phys. Acta {\bf 6}, 110 (1933);
V. C. Rubin and W. K. Ford, Astrophys. J.
{\bf 159}, 379 (1970); Y. Sofue and V. Rubin, Ann. Rev. Astron.
Astrophys. {\bf 39}, 137 (2001).

\bibitem{DeFelice:2010aj}
%\cite{Faraoni:2008mf}
%\bibitem{Faraoni:2008mf}
  V.~Faraoni,
  %``f(R) gravity: Successes and challenges,''
  arXiv:0810.2602 [gr-qc];
  %%CITATION = ARXIV:0810.2602;%%
  A.~De Felice and S.~Tsujikawa,
  %``f(R) theories,''
  Living Rev.\ Rel.\  {\bf 13}, 3 (2010);
%  [arXiv:1002.4928 [gr-qc]];
   %\cite{Sotiriou:2008rp}
%\bibitem{Sotiriou:2008rp}
  T.~P.~Sotiriou and V.~Faraoni,
  %``f(R) Theories Of Gravity,''
  Rev.\ Mod.\ Phys.\  {\bf 82}, 451 (2010);
%  [arXiv:0805.1726 [gr-qc]]
  %%CITATION = ARXIV:0805.1726;%%
  %%CITATION = ARXIV:1002.4928;%%
  T.~Clifton, P.~G.~Ferreira, A.~Padilla and C.~Skordis,
  %``Modified Gravity and Cosmology,''
  Phys.\ Rept.\  {\bf 513}, 1 (2012).
  %[arXiv:1106.2476 [astro-ph.CO]].
  %%CITATION = ARXIV:1106.2476;%%
  %\cite{DeFelice:2010aj}


%\cite{Kamenshchik:2001cp}
\bibitem{Kamenshchik:2001cp}
  A.~Y.~.Kamenshchik, U.~Moschella and V.~Pasquier,
  %``An Alternative to quintessence,''
  Phys.\ Lett.\ B {\bf 511}, 265 (2001).%  [gr-qc/0202064].
  %%CITATION = GR-QC/0202064;%%


%\cite{Sandvik:2002jz}
\bibitem{Sandvik:2002jz}
  H.~Sandvik, M.~Tegmark, M.~Zaldarriaga and I.~Waga,
  %``The end of unified dark matter?,''
  Phys.\ Rev.\ D {\bf 69}, 123524 (2004);
 % [astro-ph/0212114];
  %%CITATION = ASTRO-PH/0212114;%%
%\cite{Amendola:2003bz}
%\bibitem{Amendola:2003bz}
  L.~Amendola, F.~Finelli, C.~Burigana and D.~Carturan,
  %``WMAP and the generalized Chaplygin gas,''
  JCAP {\bf 0307}, 005 (2003);
  %[astro-ph/0304325];
  %%CITATION = ASTRO-PH/0304325;%%
  %\cite{Reis:2003mw}
%\bibitem{Reis:2003mw}
  R.~R.~R.~Reis, I.~Waga, M.~O.~Calvao and S.~E.~Joras,
  %``Entropy perturbations in quartessence Chaplygin models,''
  Phys.\ Rev.\ D {\bf 68}, 061302 (2003);
%  [astro-ph/0306004];
  %%CITATION = ASTRO-PH/0306004;%%
%\cite{Gorini:2007ta}
%\bibitem{Gorini:2007ta}
  V.~Gorini, A.~Y.~Kamenshchik, U.~Moschella, O.~F.~Piattella and A.~A.~Starobinsky,
  %``Gauge-invariant analysis of perturbations in Chaplygin gas unified models of dark matter and dark energy,''
  JCAP {\bf 0802}, 016 (2008).
  %[arXiv:0711.4242 [astro-ph]].
  %%CITATION = ARXIV:0711.4242;%%

%\cite{Banados:2008rm}

%\cite{Banados:2007qz}
\bibitem{Banados:2007qz}
  M.~Banados,
  %``The Ground-state of General Relativity and Dark Matter,''
  Class.\ Quant.\ Grav.\  {\bf 24}, 5911 (2007).
%  [hep-th/0701169].
  %%CITATION = HEP-TH/0701169;%%
\bibitem{Banados:2008rm}
  M.~Banados,
  %``A Born-Infeld action for dark energy and dark matter,''
  Phys.\ Rev.\  D {\bf 77}, 123534 (2008).
%  [arXiv:0801.4103 [hep-th]].
  %%CITATION = PHRVA,D77,123534;%%

\bibitem{Eddington} A.S. Eddington, ``The Mathematical Theory of Relativity'', Cambridge University Press (1924); E. Schr\"{o}dinger, ``Spacetime Structure'', Cambridge University Press (1950).

%\cite{Deser:1998rj}
\bibitem{Deser:1998rj} 
  S.~Deser and G.~W.~Gibbons,
  %``Born-Infeld-Einstein actions?,''
  Class.\ Quant.\ Grav.\  {\bf 15}, L35 (1998).
  %[hep-th/9803049].
  %%CITATION = HEP-TH/9803049;%%

\bibitem{Witten1} E. Witten, Nucl. Phys. B {\bf 311}, 46 (1988).







%\cite{Banados:2008jx}
\bibitem{Banados:2008jx}
  M.~Banados,
  %``Eddington-Born-Infeld action and the dark side of general relativity,''
  arXiv:0807.5088 [gr-qc].
  %%CITATION = ARXIV:0807.5088;%%


%\cite{Rodrigues:2008kv}
\bibitem{Rodrigues:2008kv}
  D.~C.~Rodrigues,
  %``Evolution of Anisotropies in Eddington-Born-Infeld Cosmology,''
  Phys.\ Rev.\  D {\bf 78}, 063013 (2008).
 % [arXiv:0806.3613 [gr-qc]].
  %%CITATION = PHRVA,D78,063013;%%


\bibitem{Skordis}
C. Skordis
%Eddington-Born-Infeld theory and the dark sector
Nucl. Phys. B (Proc. Suppl.) {\bf 194}, 338 (2009).



%\cite{Banados:2008fj}
\bibitem{Banados:2008fj}
  M.~Banados, P.~G.~Ferreira and C.~Skordis,
  %``Eddington-Born-Infeld gravity and the large scale structure of the
  %Universe,''
    Phys. Rev. D {\bf 79}, 063511 (2009);
  %arXiv:0811.1272 [astro-ph].
  %%CITATION = ARXIV:0811.1272;%%
%\cite{Banados:2008fi}
%\bibitem{Banados:2008fi}
  M.~Banados, A.~Gomberoff, D.~C.~Rodrigues and C.~Skordis,
  %``A Note on bigravity and dark matter,''
  Phys.\ Rev.\ D {\bf 79}, 063515 (2009).
 % [arXiv:0811.1270 [gr-qc]].
  %%CITATION = ARXIV:0811.1270;%%


%\cite{Banados:2010ix}
\bibitem{Banados:2010ix}
  M.~Banados and P.~G.~Ferreira,
  %``Eddington's theory of gravity and its progeny,''
  Phys.\ Rev.\ Lett.\  {\bf 105}, 011101 (2010).
 % [arXiv:1006.1769 [astro-ph.CO]].
  %%CITATION = ARXIV:1006.1769;%%

%\cite{Pani:2011mg}
\bibitem{Pani:2011mg}
  P.~Pani, V.~Cardoso and T.~Delsate,
  %``Compact stars in Eddington inspired gravity,''
  Phys.\ Rev.\ Lett.\  {\bf 107}, 031101 (2011).
 % [arXiv:1106.3569 [gr-qc]].
  %%CITATION = ARXIV:1106.3569;%%

%\cite{Delsate:2012ky}
\bibitem{Delsate:2012ky}
  T.~Delsate and J.~Steinhoff,
  %``Singular and nonsingular features of Eddington inspired Born-Infeld Gravity,''
  arXiv:1201.4989 [gr-qc].
  %%CITATION = ARXIV:1201.4989;%%


%\cite{EscamillaRivera:2012vz}
\bibitem{EscamillaRivera:2012vz} 
  C.~Escamilla-Rivera, M.~Banados and P.~G.~Ferreira,
  %``A tensor instability in the Eddington inspired Born-Infeld Theory of Gravity,''
  Phys.\ Rev.\ D {\bf 85}, 087302 (2012).
%  [arXiv:1204.1691 [gr-qc]].
  %%CITATION = ARXIV:1204.1691;%%








\bibitem{Avelino:2012ge}
  P.~P.~Avelino,
  %``Eddington-inspired Born-Infeld gravity: astrophysical and cosmological constraints,''
  arXiv:1201.2544 [astro-ph.CO].
  %%CITATION = ARXIV:1201.2544;%%


  %\cite{Pani:2012qb}
\bibitem{Pani:2012qb}
  P.~Pani, T.~Delsate and V.~Cardoso,
  %``Eddington-inspired Born-Infeld gravity. Phenomenology of non-linear gravity-matter coupling,''
  arXiv:1201.2814 [gr-qc].
  %%CITATION = ARXIV:1201.2814;%%

\bibitem{WMAP7} E.~Komatsu \textit{et al.} {[}WMAP Collaboration{]},
%``Seven-Year Wilkinson Microwave Anisotropy Probe 
%(WMAP) Observations: Cosmological Interpretation,''
Astrophys.\ J.\ Suppl.\ \ \textbf{192}, 18 (2011).

\bibitem{Percival} W.~J.~Percival \textit{et al.}, %``Baryon Acoustic Oscillations in the Sloan Digital Sky Survey Data Release 7
%Galaxy Sample,''
Mon.\ Not.\ Roy.\ Astron.\ Soc.\ \textbf{401}, 2148 (2010).

\bibitem{hicken} M.~Hicken \textit{et al.}, Astrophys.~J.~\textbf{700},
1097 (2009).


\bibitem{chisq} A.~De Felice and S.~Tsujikawa,
  %``Cosmological constraints on extended Galileon models,''
  JCAP {\bf 1203}, 025 (2012).




















\end{thebibliography}
\end{document}